\documentclass[12pt]{article}
\usepackage[dvips]{graphicx}
\usepackage{pdproc}

  \makeatletter 
  \def\@cite#1{[#1]} 
  \makeatother    
\begin{document}

\renewcommand{\thefootnote}{\alph{footnote}}
\newcommand{\dfrac}[2]{\frac{\strut \displaystyle{#1}}
                      {\strut \displaystyle{#2}}}
\newcommand{\tr}{\mbox{tr}}
\newcommand{\VEV}[1]{\langle #1 \rangle}

\title{
 Oblique Corrections in Deconstructed Higgsless Models
\footnote{Talk presented by M.K. at 
{\it SUSY 2004 : The 12th International Conference on Supersymmetry
and Unification of Fundamental Interactions}, 
held at Epochal Tsukuba, Tsukuba, Japan, 
June 17-23, 2004.
This talk is based on the work done in \cite{CSHKT}.}
}

\author{ MASAFUMI KURACHI}

\address{ 
Department of Physics, Nagoya University \\
Nagoya 464-8602, Japan
}

\author{ R. SEKHAR CHIVUKULA and ELIZABETH H. SIMMONS}

\address{ 
Department of Physics and Astronomy, Michigan State University \\
East Lansing, MI 48824, USA
}

\author{ HONG-JIAN HE}

\address{ 
Department of Physics, University of Texas \\
Austin, TX 78712, USA
}

\author{ MASAHARU TANABASHI}

\address{ 
Department of Physics, Tohoku University \\
Sendai 980-8578, Japan
}

\abstract{
In this talk, using deconstruction, we analyze the form of the
corrections to the electroweak interactions in a large class of 
``Higgsless'' models of electroweak symmetry breaking, 
allowing for arbitrary 5-D geometry, position-dependent 
gauge coupling, and brane kinetic energy terms. 
Many models considered in the literature, 
including those most likely to be
phenomenologically viable, are in this class. 
By analyzing the asymptotic behavior of
the correlation function of gauge currents at high momentum, we extract
the exact form of the relevant correlation functions at tree-level and
compute the corrections to precision electroweak observables in terms of
the spectrum of heavy vector bosons. We determine when nonoblique
corrections due to the interactions of fermions with the heavy vector
bosons become important, and specify the form such interactions can
take. In particular we find that in this class of models, so long
as the theory remains unitary, $S-4 \cos^2\theta_W T
> {\cal O}(1)$, where $S$ and $T$ are the usual oblique parameters.
}

\normalsize\baselineskip=15pt

\section{The Model and Its Relatives}

Recently, ``Higgsless'' models of electroweak symmetry breaking have
been proposed \cite{Csaki:2003dt}.  
Based on five-dimensional gauge theories compactified on an interval,
these models achieve unitarity of electroweak boson self-interactions 
through the exchange of a tower of massive vector bosons 
\cite{SekharChivukula:2001hz}.
Precision electroweak constraints \cite{Peskin:1992sw} arising from
corrections to the $W$ and $Z$ propagators in these models have
previously been investigated \cite{Csaki:2003zu}
 in the continuum and, in the case of weak coupling,
 \cite{Foadi:2003xa,Chivukula:2004kg} using deconstruction
 \cite{Arkani-Hamed:2001ca}.

\begin{figure}[htb]
\begin{center}
\includegraphics*[width=11cm]{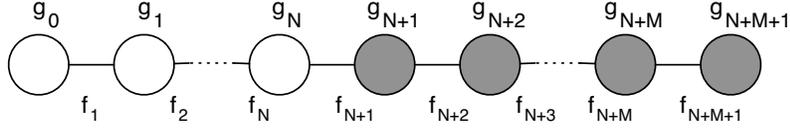}
\caption{%
Moose diagram for the class of models analyzed in this talk. $SU(2)$
 gauge groups are shown as open circles; $U(1)$ gauge groups as shaded
 circles. The fermions couple to gauge groups 0 and $N+1$.  The
 values of the gauge couplings $g_i$ and f-constants $f_i$ are
 arbitrary. 
}
\label{fig:TheMoose}
\end{center}
\end{figure}

The model we analyze, shown diagrammatically 
(using ``moose notation'' \cite{Georgi:1986hf,Arkani-Hamed:2001ca}) in
Fig. \ref{fig:TheMoose}, incorporates an 
$SU(2)^{N+1} \times U(1)^{M+1}$ gauge group, and $N+1$ 
nonlinear $(SU(2)\times SU(2))/SU(2)$ sigma models adjacent to $M$
$(U(1) \times U(1))/U(1)$ sigma models in which the global symmetry groups 
in adjacent sigma models are identified with the corresponding factors of the gauge group.
The Lagrangian for this model at $O(p^2)$ is given by
\begin{equation}
  {\cal L}_2 =
  \frac{1}{4} \sum_{j=1}^{N+M+1} f_j^2 \mbox{tr}\left(
    (D_\mu U_j)^\dagger (D^\mu U_j) \right)
  - \sum_{j=0}^{N+M+1} \dfrac{1}{2g_j^2} \mbox{tr}\left(
    F^j_{\mu\nu} F^{j\mu\nu}
    \right),
\label{lagrangian}
\end{equation}
with
\begin{equation}
  D_\mu U_j = \partial_\mu U_j - i A^{j-1}_\mu U_j 
                               + i U_j A^{j}_\mu,
\end{equation}
where all  gauge fields $A^j_\mu$ $(j=0,1,2,\cdots, N+M+1)$ are
dynamical. The first $N+1$ gauge fields ($j=0,1,\ldots, N$) correspond
to $SU(2)$ gauge groups; the other $M+1$ gauge fields ($j=N+1, N+2,
\ldots, N+M+1$) correspond to $U(1)$ gauge groups.   

The fermions in this model take their weak interactions from the $SU(2)$
group at $j=0$ and their hypercharge interactions from the $U(1)$ group
with $j=N+1$, at the interface between the $SU(2)$ and $U(1)$ groups.
The neutral and charged current couplings to the fermions are thus written as 
\begin{equation}
J^\mu_3 A^0_\mu + J^\mu_Y A^{N+1}_\mu~,\ \ 
{1 \over \sqrt{2}} J^\mu_{\pm} A^{0\mp}_\mu~. 
\label{eq:current}
\end{equation}

This model includes the model analyzed in
\cite{Foadi:2003xa,Chivukula:2004kg} and 
a deconstructed version of models in \cite{Csaki:2003zu}.
In our analysis, generalizing \cite{Chivukula:2004kg}, we leave the
gauge couplings and $f$-constants 
of the deconstructed model arbitrary and parameterize the electroweak
corrections in terms of the masses of the heavy vector bosons. We
therefore obtain results which describe arbitrary 5-D geometry,
position-dependent couplings, or brane kinetic energy terms. 


\section{The Low-Energy $\rho$ Parameter}

In discussing low-energy interactions it is conventional to write  the
neutral-current Lagrangian in terms of weak and electromagnetic currents
as 
\begin{equation}
{\cal L}_{nc} = - {1\over 2} A(Q^2)  J^\mu_3 J_{3\mu} - B(Q^2) J^\mu_3 J_{Q\mu}
- {1\over 2} C(Q^2) J^\mu_Q J_{Q\mu}~.
\end{equation}
On the other hand, the charged-current Lagrangian is written as 
\begin{equation}
{\cal L}_{cc} = - {1\over 2} [G_{CC}(Q^2)]_{WW}  J^\mu_+ J_{-\mu}~.
\label{cclagrangiani}
\end{equation}
The coefficients $A(Q^2), B(Q^2), C(Q^2)$ and $[G_{CC}(Q^2)]_{WW}$ are 
obtained by calculating the neutral and the charged gauge field
propagator matrices.
From the explicit form of these coefficients, 
we find \cite{CSHKT} that the low-energy $\rho$ parameter is identically
$1$  
in the present model:
\begin{equation}
\rho \ \equiv\ {A(Q^2=0) \over [G_{CC}(Q^2=0)]_{WW}} \ =\ 1~.
\label{eq:rho-parameter}
\end{equation}
This is an extension of the result in  \cite{Chivukula:2003wj} and is
due to our specifying that the fermion hypercharges arise from the
$U(1)$ group with $j=N+1$.
\footnote{
In \cite{CSHKT2}, the linear moose model with fermions coupling to the
$p$th $(0<p<N+1)$ $SU(2)$ gauge group and the $N+1$st $U(1)$ group was
analyzed.
We can see that $\rho=1$ is satisfied for arbitrary $p$ due to 
our specifying that the fermion hypercharges arise from the
$U(1)$ group with $j=N+1$.
}

\section{Electroweak Parameters}

From the calculation of the gauge boson propagator and corresponding
residues, we see that the size of the coupling of the heavy resonances
to fermions depends crucially on the isospin asymmetry of these states.
(See the explicit form of residues in \cite{CSHKT}.)

In the case of large isospin violation, we expect that the exchange of
heavy neutral resonances may be important and result in an extra
four-fermion contribution at low-energies proportional to 
$J_Y^{\mu}J_{Y\mu}$.  
In the case of small isospin violation, we also expect that the exchange
of heavy resonances may be important and result in an extra four-fermion 
contribution to both low-energy neutral and charged current exchange
proportional to the square of the weak $SU(2)$ currents 
$\vec{J}^{\mu}\cdot\vec{J}_{\mu}$. 
Given the constraint that the low-energy $\rho$ parameter must equal to 
one, the matrix element for four-fermion neutral and charged current
processes are characterized by 
\begin{eqnarray}
 {\cal M}_{\rm NC}
  &=& e^2 \dfrac{{\cal Q} {\cal Q}'}{Q^2}
  +\dfrac{(I_3-s^2 {\cal Q}) (I'_3 - s^2 {\cal Q}')}
  {\left(\dfrac{s^2 c^2}{e^2}-\dfrac{S}{16\pi}\right)Q^2
  +\dfrac{1}{4\sqrt{2} G_F}
  \left(1 - \alpha Te^2 + \dfrac{\alpha \delta}{4 s^2 c^2}\right)
  }\nonumber\\
  & &+ \sqrt{2} G_F \,{\alpha \delta\over s^2 c^2}\, I_3 I'_3
   - 4\sqrt{2} G_F \alpha T (I_3-s^2 {\cal Q}) (I'_3 - s^2 {\cal Q}')~, 
  \label{eq:NC3}
\end{eqnarray}
\begin{eqnarray}
 {\cal M}_{\rm CC}
  =  \dfrac{(I_{+} I'_{-} + I_{-} I'_{+})/2}
  {\left(\dfrac{s^2}{e^2}-\dfrac{S+U}{16\pi}\right)Q^2
  +\dfrac{1}{4\sqrt{2} G_F}
  \left(1 + \dfrac{\alpha \delta}{4 s^2 c^2}\right)
  }
  + \sqrt{2} G_F\, {\alpha \delta\over s^2 c^2} \, 
  {(I_{+} I'_{-} + I_{-} I'_{+}) \over 2}~. 
  \label{eq:CC3}
\end{eqnarray}
Here $I^{(\prime)}_a$ and ${\cal Q}^{(\prime)}$ are weak isospin and
charge of the corresponding fermion, 
$\alpha = e^2/4\pi$, $G_F$ is the usual Fermi constant, 
and the weak mixing angle (as defined by the on-shell $Z$ coupling) is
denoted by $s^2$ ($c^2\equiv 1-s^2$).  
From these analysis, we can see when nonoblique corrections become
important, and we can estimate the size of them.

We can obtain expressions for $S$, $T$, and $U$ by comparing the
residues as calculated from (\ref{eq:NC3}), (\ref{eq:CC3}) with
their values as calculated in terms of the gauge boson spectrum.
Independent of the amount of isospin violation, we have found the
following constraint on the oblique electroweak parameters $S$ and $T$ 
\begin{equation}
\alpha S - 4 \cos^2\theta_W\,\alpha T = 4 \sin^2\theta_W \cos^2\theta_W M^2_Z\,   
\sum_{\hat{n}=1}^N {1\over M^2_{W\hat{n}}} ~,
\label{eq:summary}
\end{equation}
where $M_{W\hat{n}}$ is the mass of the $n$th heavy charged resonances.
In any unitary theory \cite{SekharChivukula:2001hz}, we expect the mass
of the lightest vector $M_{W\hat{1}}$ to be less than $\sqrt{8 \pi} v$
($v\approx 246$ GeV). 
Evaluating eqn.(\ref{eq:summary}) we see that we expect 
$S-4 \cos^2\theta_W T$ to be of order one-half  or larger, generalizing
the result of \cite{Chivukula:2004kg}.

\section{Acknowledgements}

M.K. acknowledges support by the 21st Century COE Program of Nagoya
University provided by JSPS (15COEG01). M.T.'s work is supported in part
by the JSPS Grant-in-Aid for Scientific Research No.16540226. H.J.H. is 
supported by the US Department of Energy grant 
DE-FG03-93ER40757.

\bibliographystyle{plain}

\begin{thebibliography}{99}
\bibitem{CSHKT}
R.~S. Chivukula, E.~H. Simmons, H.-J. He, M. Kurachi, and
M. Tanabashi, 
hep-ph/0406077.

\bibitem{Csaki:2003dt}
C.~Csaki, C.~Grojean, H.~Murayama, L.~Pilo, and J.~Terning, hep-ph/0305237.

\bibitem{SekharChivukula:2001hz}
R.~Sekhar~Chivukula, D.~A. Dicus, and H.-J. He, 
  {\it Phys. Lett.} {\bf B525} (2002)
  175--182;\ 
R.~S. Chivukula and H.-J. He,  {\it Phys. Lett.} {\bf B532} (2002)
	121--128;\ 
R.~S. Chivukula, D.~A. Dicus, H.-J. He, and S.~Nandi,
  {\it Phys. Lett.} {\bf B562} (2003)
  109--117.

\bibitem{Peskin:1992sw}
M.~E. Peskin and T.~Takeuchi, {\it Phys. Rev.} {\bf D46} (1992) 381--409.

\bibitem{Csaki:2003zu}
C.~Csaki, C.~Grojean, L.~Pilo, and J.~Terning, {\it Phys. Rev. Lett.} {\bf 92}
  (2004) 101802;\ 
G.~Cacciapaglia, C.~Csaki, C.~Grojean, and J.~Terning, hep-ph/0401160;\ 
Y.~Nomura, {\it JHEP} {\bf 11} (2003) 050;\ 
R.~Barbieri, A.~Pomarol, and R.~Rattazzi, hep-ph/0310285;\ 
H.~Davoudiasl, J.~L. Hewett, B.~Lillie, and T.~G. Rizzo,
	hep-ph/0312193;\ 
G.~Burdman and Y.~Nomura, hep-ph/0312247;\ 
H.~Davoudiasl, J.~L. Hewett, B.~Lillie, and T.~G. Rizzo,
  {\it JHEP} {\bf 05} (2004) 015;\ 
R.~Barbieri, A.~Pomarol, R.~Rattazzi, and A.~Strumia, hep-ph/0405040.

\bibitem{Foadi:2003xa}
R.~Foadi, S.~Gopalakrishna, and C.~Schmidt, {\it JHEP} {\bf 03} (2004) 042.

\bibitem{Chivukula:2004kg}
R.~S. Chivukula, M.~Kurachi, and M.~Tanabashi, {\it JHEP} {\bf 06} (2004) 004.

\bibitem{Arkani-Hamed:2001ca}
N.~Arkani-Hamed, A.~G. Cohen, and H.~Georgi, 
   {\it Phys. Rev. Lett.} {\bf 86} (2001) 4757--4761;\ 
C.~T. Hill, S.~Pokorski, and J.~Wang, {\it Phys. Rev.} {\bf D64} (2001)
	105005. 

\bibitem{Georgi:1986hf}
H.~Georgi, {\it Nucl. Phys.} {\bf B266} (1986) 274.

\bibitem{Chivukula:2003wj}
R.~S. Chivukula, H.-J. He, J.~Howard, and E.~H. Simmons, 
  {\it Phys. Rev.}
  {\bf D69} (2004) 015009.

\bibitem{CSHKT2}
R.~S. Chivukula, E.~H. Simmons, H.-J. He, M. Kurachi, and
M. Tanabashi, 
hep-ph/0408262.
\end{thebibliography}

\end{document}